\pdfoutput=1

\documentclass[11pt,twoside,a4paper,cmspaper,final,collab]{cms-tdr}
\def\svnVersion{414488}\def\cmsCernNoTag{CERN-EP/2016-248}\def\cmsCernDate{\today}\def\cmsMessage{Published in Physics Letters B as \href{http://dx.doi.org/10.1016/j.physletb.2017.04.031}{\doi{10.1016/j.physletb.2017.04.031.}}}

\begin{document}\cmsNoteHeader{HIN-15-001}

\hyphenation{had-ron-i-za-tion}
\hyphenation{cal-or-i-me-ter}
\hyphenation{de-vices}
\RCS$Revision: 396073 $
\RCS$HeadURL: svn+ssh://svn.cern.ch/reps/tdr2/papers/HIN-15-001/trunk/HIN-15-001.tex $
\RCS$Id: HIN-15-001.tex 396073 2017-03-26 18:58:08Z alverson $

\newlength\cmsFigWidth
\ifthenelse{\boolean{cms@external}}{\setlength\cmsFigWidth{0.49\textwidth}}{\setlength\cmsFigWidth{0.6\textwidth}}
\ifthenelse{\boolean{cms@external}}{\providecommand{\cmsLeft}{top\xspace}}{\providecommand{\cmsLeft}{left\xspace}}
\ifthenelse{\boolean{cms@external}}{\providecommand{\cmsRight}{bottom\xspace}}{\providecommand{\cmsRight}{right\xspace}}

\newcommand {\npart}  {\ensuremath{N_{\text{part}}}\xspace}
\newcommand {\ncoll}  {\ensuremath{N_{\text{coll}}}\xspace}
\newcommand{\raa}{\ensuremath{R_\mathrm{AA}}\xspace}
\newcommand{\taa}{\ensuremath{T_\mathrm{AA}}\xspace}
\newcommand{\sqrts}{\ensuremath{\sqrt{s}}\xspace}
\newcommand{\sqrtsnn}{\ensuremath{\sqrt{\smash[b]{s_{_{\mathrm{NN}}}}}}\xspace}
\newcommand{\QQbar}{\ensuremath{\mathrm{Q}\ensuremath{\overline{\mathrm{Q}}}}\xspace}
\newcommand{\HYDJET} {{\textsc{hydjet}}\xspace}
\cmsNoteHeader{HIN-15-001}
\title{Suppression of $\PgU$(1S), $\PgU$(2S), and $\PgU$(3S) quarkonium states in PbPb  collisions at $\sqrtsnn = 2.76$\TeV}

\date{\today}

\abstract{
The production yields of $\PgU$(1S), $\PgU$(2S), and $\PgU$(3S) quarkonium states are measured through their decays into muon pairs in the CMS detector, in PbPb and pp collisions at the centre-of-mass energy per nucleon pair of 2.76\TeV. The data correspond to integrated luminosities of 166\mubinv and 5.4\pbinv for PbPb and pp collisions, respectively. Differential production cross sections are reported as functions of $\PgU$ rapidity $y$ up to 2.4, and transverse momentum \pt up to 20\GeVc. A strong centrality-dependent suppression is observed in PbPb relative to pp collisions, by factors of up to ${\approx}2$ and 8, for the $\PgU$(1S) and $\PgU$(2S) states, respectively. No significant dependence of this suppression is observed as a function of $y$ or \pt. The $\PgU$(3S) state is not observed in PbPb collisions, which corresponds to a suppression for the centrality-integrated data by at least a factor of ${\approx}7$ at a 95\% confidence level. The observed suppression is in agreement with theoretical scenarios modeling the sequential melting of quarkonium states in a quark gluon plasma.
}

\hypersetup{%
pdfauthor={CMS Collaboration},%
pdftitle={Suppression of Upsilon(1S), Upsilon(2S) and Upsilon(3S) production in PbPb  collisions at sqrt(s[NN]) = 2.76 TeV},%
pdfsubject={CMS},%
pdfkeywords={CMS, physics, heavy ions, dimuons, bottomonia}}

\maketitle

\hyphenation{col-la-bo-ra-tion}
\hyphenation{vo-lume}
\hyphenation{mid-rapidity}

\section{Introduction} \label{sec:intro}

At large energy density and high temperature, strongly interacting matter is predicted by lattice QCD calculations to consist of a deconfined system of quarks and gluons~\cite{Karsch:2003jg}. This state, often referred to as ``quark gluon plasma'' (QGP)~\cite{Shuryak:1977ut}, constitutes the main object of studies using high energy heavy ion collisions.

The formation of QGP in nuclear collisions is studied in a variety of ways. One of its most striking signatures is the sequential suppression of quarkonium states, both in the charmonium (\JPsi, $\psi'$, $\chi_\mathrm{c}$, etc.) and the bottomonium ($\PgU\mathrm{(1S,\,2S,\,3S)}$, $\chi_\mathrm{b}$, etc.) families. Historically, this phenomenon was proposed as direct evidence of deconfinement because, in the deconfined medium, the binding potential between the constituents of a quarkonium state, a heavy quark and its antiquark (\QQbar), should be screened by the colour charges of the surrounding light quarks and gluons~\cite{Matsui:1986dk, Digal:2001ue}. The suppression of quarkonium production is predicted to occur above the critical temperature of the medium ($T_\mathrm{c}$) and to depend on the \QQbar~binding energy. Since the \PgUa\ is the most tightly bound state among all quarkonia, it is expected to have the highest dissociation temperature. Estimates of dissociation temperatures are given in Ref.~\cite{Mocsy:2007jz}: $T_{\text{dissoc}} \approx 2 \,T_\mathrm{c}$, $1.2\,T_\mathrm{c},$ and $1\,T_\mathrm{c}$ for the \PgUa, \PgUb, and \PgUc\ states, respectively. Other medium effects, such as regeneration from initially uncorrelated quark-antiquark pairs~\cite{Thews:2000rj,Andronic:2006ky} or absorption by comoving particles~\cite{Gavin:1996yd,Capella:1996va} can modify quarkonium production in heavy ion collisions. Furthermore, nuclear effects such as modifications of parton distributions inside nuclei~\cite{Vogt:2010aa} or energy loss processes in nuclear matter~\cite{Arleo:2014oha} are expected to affect the production of quarkonia independently of any QGP formation. An admixture of several of the above-mentioned effects in the context of bottomonium production is investigated in Refs.~\cite{Emerick:2011xu,Strickland:2011aa} and a recent review on quarkonium production can be found in Ref.~\cite{Andronic:2015wma}. 

The suppression of  \PgUa\ production in heavy ion collisions relative to pp yields scaled by the number of binary nucleon-nucleon (NN) collisions was first measured by CMS~\cite{Chatrchyan:2012np} in the midrapidity range $\abs{y}< 2.4$, then by ALICE at forward rapidities $2.5 < y < 4$~\cite{Abelev:2014nua}. Both measurements were done at the CERN LHC in PbPb collisions at a centre-of-mass energy per nucleon pair, \sqrtsnn, of 2.76\TeV. A larger suppression of the \PgUb\ and \PgUc\ was first suggested~\cite{Chatrchyan:2011pe} then observed~\cite{Chatrchyan:2012lxa} by CMS. In pPb collisions at $\sqrtsnn = 5.02$\TeV, ALICE~\cite{Abelev:2014oea} and LHCb~\cite{Aaij:2014mza} reported \PgUa\ yields that are slightly suppressed along the
p-going forward direction, possibly indicating the importance of nuclear effects. Lacking pp reference data at $\sqrtsnn=5.02$\TeV, the pp yields were estimated by interpolating results at 2.76, 7, and 8\TeV~\cite{Abelev:2014oea}, or by scaling data at 8\TeV~\cite{Aaij:2014mza}. The \PgUb\ and \PgUc\ were reported by CMS to be slightly more suppressed than the \PgUa\ ground state in pPb collisions~\cite{Chatrchyan:2013nza}. 
At the BNL RHIC, STAR reported no significant suppression of the overlapping $\PgU$(1S+2S+3S) states in dAu collisions at $\sqrtsnn = 200$\GeV, while observing a suppression in central AuAu collisions at the same energy~\cite{Adamczyk:2013poh}. Altogether, these results are interpreted as a sequential suppression of the three states in nucleus-nucleus collisions~\cite{Emerick:2011xu,Strickland:2011aa}, with the tighter bound states disappearing less in the QGP.

This Letter reports the production yields of \PgUa, \PgUb, and \PgUc\ for PbPb and pp data at the same $\sqrtsnn = 2.76$\TeV, using integrated luminosities of 166\mubinv and 5.4\pbinv, respectively. The two sets of data correspond to approximately the same number of NN collisions. 
The pp sample collected in 2013 contains 20 times more events than the 2011 data used previously~\cite{Chatrchyan:2012np, Chatrchyan:2011pe, Chatrchyan:2012lxa}, allowing further differential studies with respect to the \PgU\ meson rapidity and transverse momentum. Muon reconstruction is improved in PbPb collisions relative to Ref.~\cite{Chatrchyan:2012lxa}, yielding a 35\% increase in the number of measured \PgU\ candidates. In total, the improved reconstruction and a relaxed muon-\pt selection provide almost twice the number of \PgUa\ candidates used in Ref.~\cite{Chatrchyan:2012lxa}. The yields in PbPb and pp events are used to extract nuclear modification factors, $R_\mathrm{AA}$. 

\section{The CMS detector} \label{sec:detector}

A detailed description of the CMS detector, together with a definition of the coordinate system used and the relevant kinematic variables, can be found in Ref.~\cite{Adolphi:2008zzk}. The central feature of the CMS apparatus is a superconducting solenoid of 6\unit{m} internal diameter. A silicon tracker, a crystal electromagnetic
calorimeter, and a brass and scintillator hadron calorimeter reside within the magnetic field volume.

Muons are detected in the pseudorapidity interval $\abs{\eta}< 2.4$ using gas-ionization detectors made of three technologies: drift tubes, cathode strip chambers, and resistive-plate chambers, embedded in the steel flux-return yoke of the solenoid. The silicon tracker is composed of pixel detectors (three barrel layers
and two forward disks on either side of the detector, made of 66~million $100{\times}150\mum^2$ pixels) followed by microstrip detectors (ten barrel layers, and three inner and nine forward disks on either side of the detector, with strips of pitch between 80 and 180\mum). The transverse momentum of muons matched to
tracks reconstructed in the silicon detector is measured with a resolution better than $1.5$\% for \pt values smaller than 100\GeVc~\cite{TRK-10-004}. This high resolution is the result of the 3.8\unit{T} magnetic field and the high granularity of the silicon tracker.

In addition, CMS has extensive forward calorimetry, including two steel and quartz-fibre Che\-ren\-kov hadron forward (HF) calorimeters, that cover the range $2.9 < \abs{\eta}< 5.2$. These detectors are used in the present analysis to select events and to determine the centrality of PbPb collisions, as described in the next
section. 

\section{Data selections}

\subsection{Event selection and centrality} \label{sec:centrality}

To select purely inelastic hadronic PbPb collisions, contributions from ultraperipheral collisions and noncollision beam backgrounds are removed, as described in Ref.~\cite{Chatrchyan:2011sx}. Events are preselected if they contain a primary vertex built from at least two tracks, and at least three signals (one in the case of pp collisions) in HF towers on each side of the interaction point with deposited energies of at least 3\GeV\ in each tower. To further suppress beam-gas events, the distribution of hits in the pixel detector along the beam direction is required to be compatible with particles originating from the event vertex. These criteria select $(97\pm3)$\% of the inelastic hadronic PbPb collisions~\cite{Chatrchyan:2011sx}, yielding an efficiency-corrected number of minimum bias (MB) events $N_\mathrm{MB}=(1.16\pm0.04)\times10^9$ for the MB sample corresponding to this analysis. The pp data correspond to an integrated luminosity of 5.4\pbinv, known to an accuracy of 3.7\% coming from the uncertainty in the calibration based on a van der Meer scan~\cite{CMS-PAS-LUM-13-002}. 

The measurements are based on events that were first selected by the \Lone trigger, a hardware-based system that uses information from the muon detectors and calorimeters. The presence of at least two muons was required, with no selection applied on their momenta. The events were then further filtered using a software-based high-level trigger, and rejected if muons were poorly reconstructed, hence likely to be misidentified. The pp and PbPb data were collected using the same trigger logic. 

The centrality of PbPb collisions is defined as the fraction of the total number of inelastic hadronic collisions, with 0\% representing collisions with the largest overlap of the two nuclei. This fraction is determined from the distribution of total energy in both HF calorimeters. Variables related to the centrality, such as the number of nucleons participating in the collision (\npart) and the nuclear overlap function (\taa)~\cite{Miller:2007ri}, are estimated using a Glauber model simulation described in Ref.~\cite{Chatrchyan:2011sx}. The value of \taa at a given centrality is equal to the number of binary NN collisions divided by the NN cross section and can be interpreted as the NN-equivalent integrated luminosity per heavy ion collision.

It is to be noted that the PbPb hadronic cross section ($7.65 \pm0.42\unit{b}$) computed with this Glauber simulation corresponds to an integrated luminosity of $152\pm9$\mubinv, compatible within 1.2 sigma with the experimental value of $166\pm8$\mubinv based on the van der Meer scan. The mean values of \taa and \npart are presented in Table~\ref{tab:glauber} for the narrow centrality bins used in the \PgUa\ analysis, the wider bins used in the \PgUb\ analysis, and the centrality-integrated estimate. The most peripheral bins are rather wide and, since quarkonium yields scale with the number of nucleon-nucleon collisions, most bottomonia are produced close to the most central edge of the bins, namely 70\% and 50\%. The $\langle\npart\rangle$ values shown in the following figures and reported in Table~\ref{tab:glauber} are computed by averaging over all MB events in a given centrality bin, and are therefore not corrected for any bias introduced by requiring the presence of the \PgU.
Also presented is the root-mean-square (RMS) of the \npart distribution in each bin. The uncertainty on \taa\ is computed by varying the Glauber parameters and the event selection inefficiency, as described in Ref.~\cite{Chatrchyan:2011sx}. In this Letter, $\langle\npart\rangle$ is used to show the centrality dependence of the measurements, while \taa directly enters into the nuclear modification factor calculation: \raa~$= N_\mathrm{PbPb} / \left( \taa \, \sigma_{\Pp\Pp}\right)$ where $N_\mathrm{PbPb}$ is the number of $\PgU$ produced per PbPb collision in a given kinematic range and $\sigma_{\rm pp}$ the corresponding $\PgU$ cross section in pp collisions.

\begin{table}[htbp]
  \begin{center}
    \topcaption{Average values of the number of participating nucleons (\npart, with the root-mean-square of its distribution in each bin), and nuclear overlap function (\taa, with its systematic uncertainty) for the centrality bins used in the \PgUa (upper rows) and \PgUb (middle) analyses. The centrality-integrated values are given in the last row.}
    \label{tab:glauber}
    \newcolumntype{x}{D{,}{\text{--}}{2.3}}
    \newcolumntype{y}{D{,}{\,}{-4.6}}            
    \newcolumntype{z}{D{,}{\,\pm\,}{5.5}}        
    \begin{tabular}{xyz}
       \hline
       \multicolumn{1}{c}{Centrality (\%)} & \multicolumn{1}{c}{$\langle \npart \rangle$ (RMS)} & \multicolumn{1}{c}{$\langle \taa \rangle$ (mb$^{-1}$)}\\\hline
      0,5	& 381,(19)  & 25.9,1.1 \\
      5,10	& 329,(22) & 20.5,0.9 \\
      10,20	& 261,(30) & 14.5,0.8 \\
      20,30	& 187,(23) & 8.80,0.58 \\
      30,40	& 130,(18) & 5.09,0.43 \\
      40,50	& 86.2,(13.6) & 2.75,0.30 \\ 
      50,70 & 42.0,(14.4) & 0.98,0.14 \\
      70,100 & 8.8,(6.0) & 0.125,0.023 \\ \hline
      0,10  & 355,(33) & 23.2,1.0 \\
      10,30 & 224,(46) & 11.6,0.7 \\
      30,50 & 108,(27) & 3.92,0.37 \\
      50,100 & 22.1,(19.3) & 0.47,0.07 \\ \hline
      0,100	& 113,(115) & 5.67,0.32 \\ \hline
    \end{tabular}
  \end{center}
\end{table}

\subsection{Muon selection}

Muons are reconstructed using a global fit to a track in the muon detectors that is matched to a track in the silicon tracker. The offline muon reconstruction algorithm used for the PbPb data has been improved relative to that used previously~\cite{Chatrchyan:2012lxa}. The efficiency has been increased by running multiple iterations in the pattern recognition step, raising the number of reconstructed \PgUa\ candidates by approximately 35\%. Background muons from cosmic rays and heavy-quark semileptonic decays are rejected by imposing a set of selection criteria on each muon track. These criteria are based on previous studies of the performance of the muon reconstruction algorithm~\cite{Chatrchyan:2012xi}. The track is required to have a hit in at least one pixel detector layer, and a respective transverse (longitudinal) distance of closest approach of less than 3\,(15)\cm from the measured primary vertex, primarily to reject cosmic ray muons and muons from hadron decays in flight. To ensure a good \pt measurement, more than 10 hits are requested in the tracker, and the $\chi^2$ per number of degrees of freedom of the trajectory fits is limited to be smaller than 10 when using the silicon tracker and the muon detectors, and smaller than 4 when using only the tracker. Pairs of oppositely charged muons are considered when the $\chi^2$ fit probability of the tracks originating from a common vertex exceeds 1\%.

\begin{figure}[hbtp]
\centering
\includegraphics[width=0.45\textwidth]{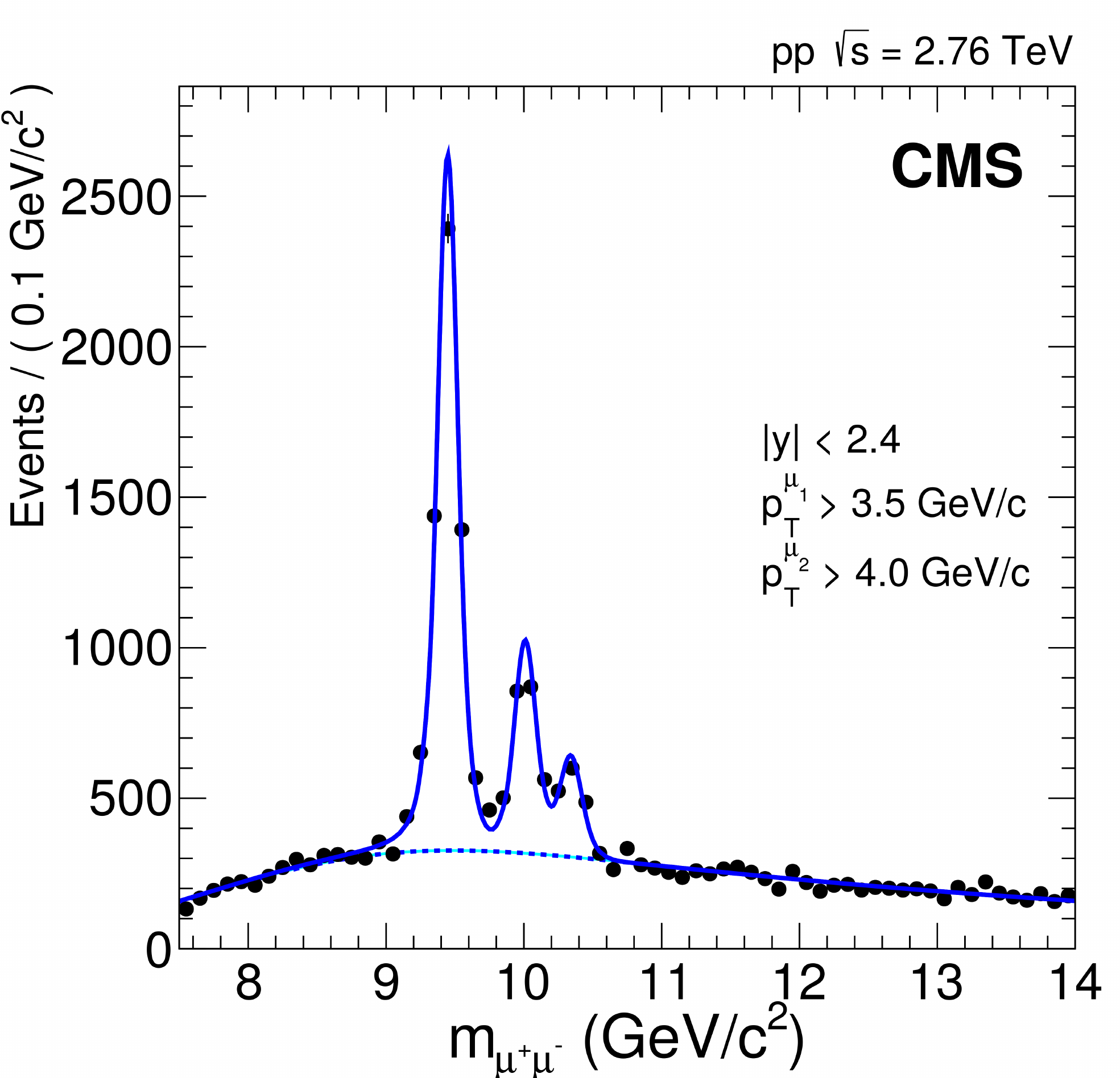}
\includegraphics[width=0.45\textwidth]{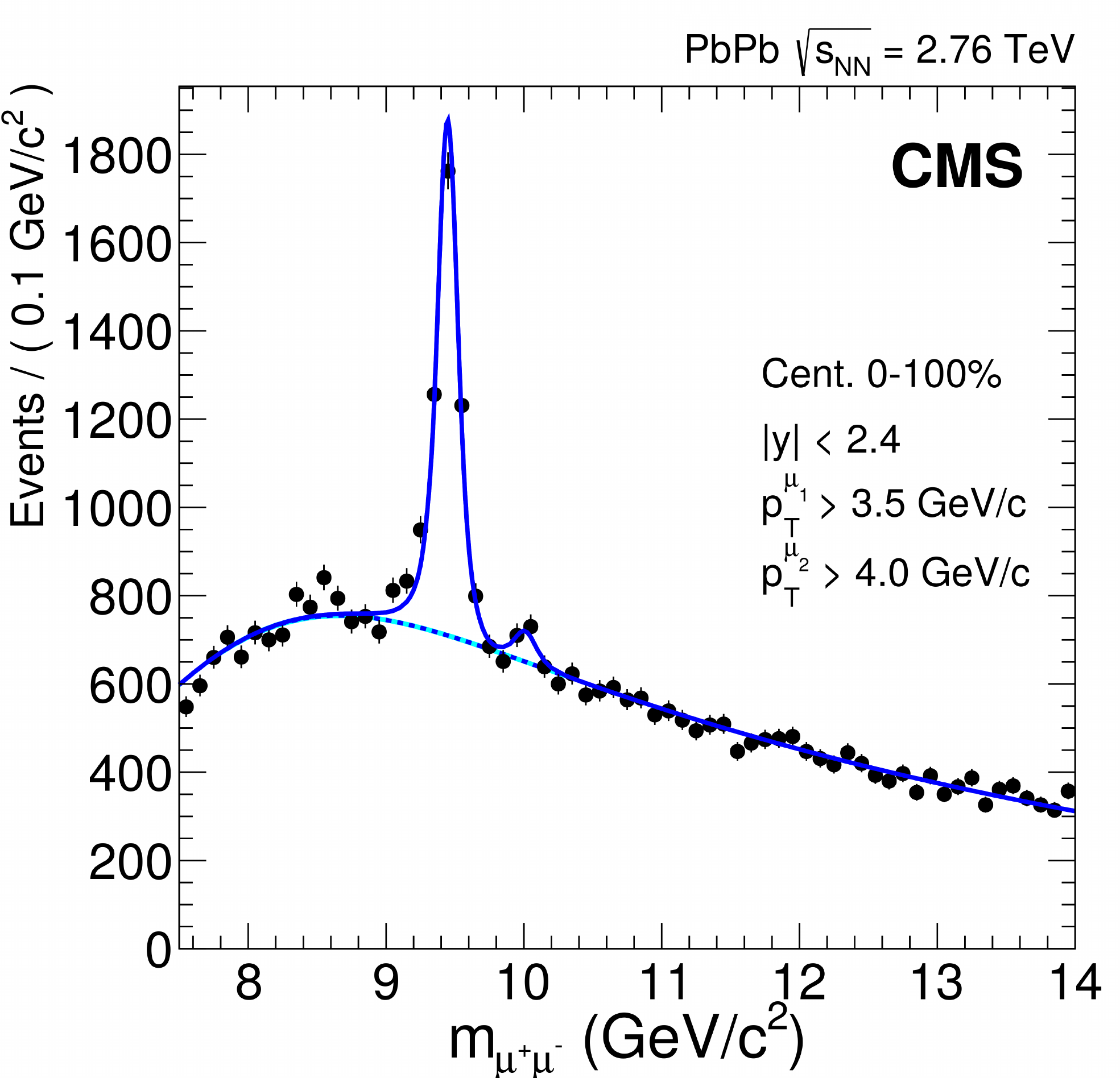}
\caption{Dimuon invariant mass distributions in pp (\cmsLeft) and centrality-integrated PbPb (\cmsRight) data at $\sqrtsnn = 2.76$\TeV, for muon pairs having
  one \pt greater than 4\GeVc\ and the other greater than 3.5\GeVc. The solid (signal + background) and dashed (background only) lines show the result of fits described in the text. 
}
\label{fig:mass} 
\end{figure}

For the \PgUb\ and \PgUc\ analyses, the transverse momentum of each muon ($\pt^{\mu}$) is required to be above 4\GeVc, as in previous publications~\cite{Chatrchyan:2012np,Chatrchyan:2011pe,Chatrchyan:2012lxa}, while one of them is relaxed down to 3.5\GeVc\ for the \PgUa\ analysis. Reducing this \pt threshold raises the \PgUa\ yield by approximately 40\%, and its statistical significance by up to 50\%, depending on the \pt and $y$ of the dimuon system. Relaxing the criterion on the second muon was also considered then discarded, since it did not significantly raise the acceptance for the \PgU\ states. The resulting invariant mass distributions are shown on Fig.~\ref{fig:mass} for the entire pp and PbPb data samples.

\section{Analysis}

\subsection{Signal extraction}

To extract the \PgUa, \PgUb, and \PgUc\ meson yields, unbinned maximum likelihood fits to the $\mu^+ \mu^-$ invariant mass spectra are performed between 7.5 and 14\GeVcc. The results for the \pt-, $y$- and centrality-integrated case are displayed as solid lines on Fig.~\ref{fig:mass}. Each \PgU\ resonance is modelled by the sum of two Crystal Ball (CB) functions~\cite{SLAC-R-236} with common mean but different widths to account for the pseudorapidity dependence of the muon momentum resolution. The CB functions are Gaussian resolution functions with the low-side tail replaced by a power law describing final-state radiation. This choice was guided by simulation studies, as well as analyses of large pp event samples collected at $\sqrts = 7$\TeV~\cite{Chatrchyan:2013yna}. Given the relatively large statistical uncertainties, the only signal model parameters that are left free in the fit are the mean of the \PgUa\ peak, and the \PgUa, \PgUb\ and \PgUc\ meson yields. The other parameters, such as the width of the \PgUa\ peak are fixed in every bin to the corresponding value obtained from simulations. The mean and width of the CB functions describing the~\PgUb\ and \PgUc\ peaks are set by the fitted \PgUa\ peak mean and the fixed \PgUa\ width, respectively, multiplied by the world-average mass ratio~\cite{Agashe:2014kda}. The parameters describing the tail of the CB function are fixed to values obtained from simulations, kept common in the three \PgU\ states, then allowed to vary when computing the associated systematic uncertainties. The background distribution is modelled by an exponential function multiplied by an error function (the integral of a Gaussian) describing the low-mass turn-on, with all parameters left free in the fit.  

With one muon having \pt greater than 4\GeVc and the other greater than 3.5\GeVc, this fitting procedure results in \PgUa\ meson yields and statistical uncertainties of $2534 \pm 76$ and $5014 \pm 87$ in centrality-integrated PbPb and pp collisions, respectively. With both muons' transverse momenta above 4\GeVc, it yields $173 \pm 41$ for \PgUb\ and $7 \pm 38$ for \PgUc\ (hence unobserved) in PbPb collisions, and $1214 \pm 51$ for \PgUb\ and $618 \pm 44$ for \PgUc\ states in pp collisions. 

\subsection{Acceptance and efficiency}

To correct yields for acceptance and efficiency in the two data samples, the three \PgU\ states have been simulated
using the \PYTHIA 6.412 generator~\cite{Sjostrand:2006za} and embedded in PbPb events simulated with \HYDJET~1.8~\cite{Lokhtin:2005px}, producing Monte Carlo (MC) events with the same settings as in Ref.~\cite{Chatrchyan:2012lxa}, including radiative tails handled by~\PHOTOS~\cite{Barberio:1990ms}. Acceptance is defined as the fraction of \PgU\ in the $\abs{y} < 2.4$ range that decay into two muons, each with $\abs{\eta^{\mu}} < 2.4$, and $\pt^{\mu_2} >$ 4\GeVc\ and $\pt^{\mu_1} >$ 3.5 or 4\GeVc\ for the \PgUa\ and \PgUb/\PgUc\ states, respectively. For the \PgUa\ state, the acceptance over the analyzed phase space averages to 35\%. For all three \PgU\ states, the acceptance is constant over most of the rapidity range, with a drop at large $\abs{y}$. When the \PgU\ meson has \pt\ ${\approx}5$\GeVc, the lower \pt\ decay muon often falls below the required momentum to reach the muon detector, resulting in a drop in acceptance for intermediate \pt.
For \PgUb\ and \PgUc\ states, where \pt for both muons is required to be above 4\GeVc, the acceptance is 28 and 33\%, respectively. Within this acceptance, the average reconstruction and trigger efficiencies are 68, 74 and 75\% for the \PgUa, \PgUb, and \PgUc\ states, respectively. The slightly lower efficiency for the \PgUa\ state arises from including lower-\pt muons, which have smaller reconstruction efficiencies, in particular at midrapidity.

The individual components of the efficiency are crosschecked using collision data and muons from \JPsi meson decays, with a technique called tag-and-probe, similar to the one described in Ref.~\cite{Chatrchyan:2013yna}. The method consists of fitting the \JPsi candidates in data and MC samples, with and without applying the probed selection criterion on one of the
muons. The muon reconstruction, identification, and trigger efficiencies in the muon detectors are probed by testing the selection response in a sample collected with single-muon triggers. The small discrepancies observed between the results for data and simulation are used to determine \pt- and $\eta$-dependent single-muon correction factors that are applied to muons in the simulation.  The net correction factors to the \PgU\ meson yields range from 3 to 18\%, the largest being located at low \pt or at large $\abs{y}$. The tracker efficiency, larger than 99\%, is also evaluated with this method by checking the presence of a track for muons that are primarily reconstructed in the muon detectors. The corresponding uncertainty is evaluated to be 0.3 and 0.6\% for each muon, for the pp and PbPb data, respectively. 

\subsection{Systematic uncertainties}

The uncertainty from the fitting procedure is estimated by performing seven changes in the fitting functions. Five of them consist of releasing one by one the originally fixed signal-shape parameters, to accommodate for possible imperfections in the simulation. The other two changes consist of adding to the default background function a first- or second-order Chebychev polynomial. The maxima of the five signal and of the two background variations are summed in quadrature, yielding systematic uncertainties from 4 to 25\% in the PbPb data and from 1 to 10\% in the pp data, for the \PgUa\ meson yield. For the less significant \PgUb\ signal, the uncertainties range from 13 to 71\% in PbPb, and from 1 to 15\% in pp data. 

The systematic uncertainty from the acceptance and efficiency estimation includes changes of the generated \pt and $y$ spectra, as well as variations of the distribution of \PgU\ candidates across event centrality, within limits imposed by the data themselves. These are propagated into bin-by-bin systematic uncertainties of 0.7 and 1.1\% on average, in pp and PbPb collisions, respectively.

Single-muon efficiencies obtained from the tag-and-probe method are assigned a systematic uncertainty from varying requirements for the tag selection, the dimuon mass range, and the distributions of the invariant mass peak and the underlying backgrounds. The maximum deviation in each $\pt^{\mu}$ and $\eta^{\Pgm}$ interval is retained as the
systematic uncertainty on the single-muon correction factors. Next, the single-muon correction factors are changed within their statistical uncertainties derived from data. To do so, one hundred variations of the single-muon efficiencies are computed, resulting in one hundred dimuon efficiency correction factors in each analysis bin. The RMS of the resulting
efficiencies, summed in quadrature with the systematic uncertainty in the efficiency correction factors, represent the overall uncertainty in muon efficiency. The resulting systematic uncertainties range from 3.2 to 7.7\% from midrapidity in pp collisions to the most forward bins in PbPb collisions. In addition, the uncertainty in the tracking efficiency of 0.3 and 0.6\% for each track is considered as fully correlated and thus doubled for dimuon candidates, and taken as a global uncertainty (common to all points).   

The relative uncertainties in the integrated luminosity of pp data (3.7\%) or the number of PbPb MB events (3\%) are also considered as global uncertainties. The uncertainties in the \taa values are given in Table~\ref{tab:glauber}.

\section{Results} \label{sec:results}

\subsection{Cross sections}

\begin{figure}[hbtp]
\centering
\includegraphics[width=0.49\textwidth]{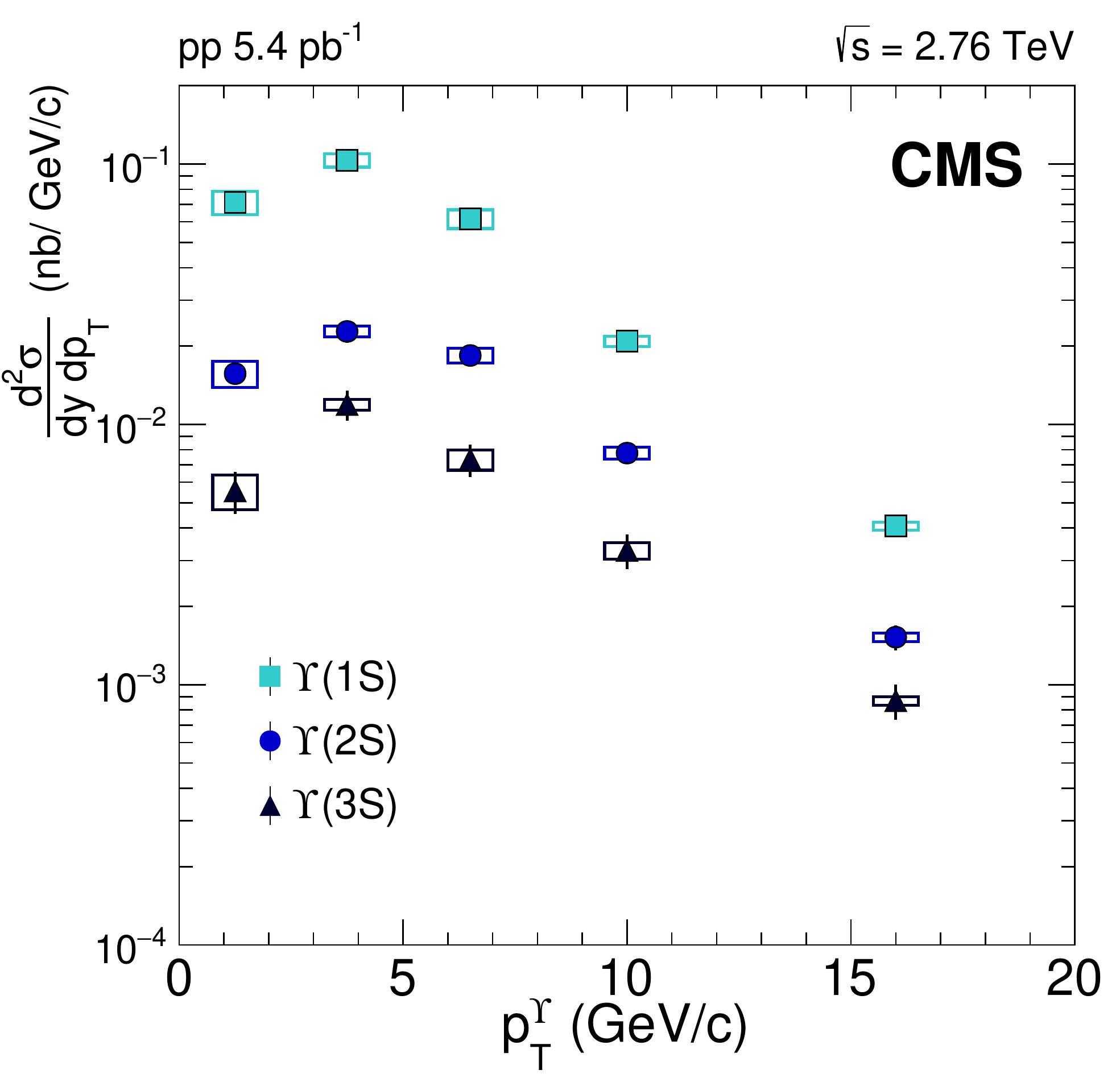}
\includegraphics[width=0.49\textwidth]{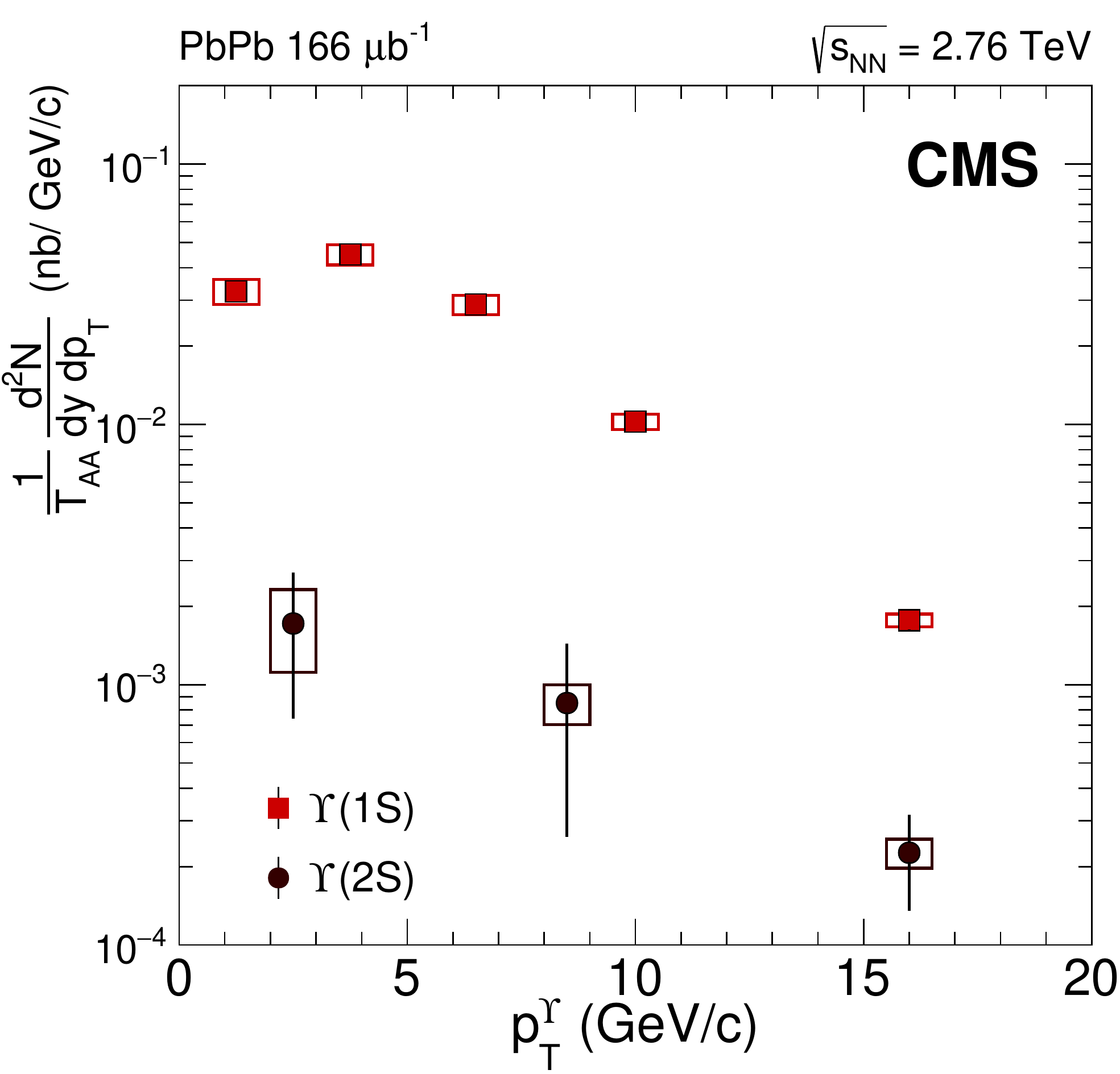}
\caption{Differential cross section for \PgU\ states as a function of their transverse momentum and per unit of rapidity in pp (\cmsLeft) and PbPb (\cmsRight) collisions. The PbPb results are integrated over centrality and divided by the number of elementary NN collisions. Statistical (systematic) uncertainties are displayed as error bars (boxes). Global relative uncertainties of 3.7\% (pp) and 6.5\% (PbPb) are not displayed.
}
\label{fig:CrossSection_ppAA_pt} 
\end{figure}

\begin{figure}[hbtp]
\centering
\includegraphics[width=0.49\textwidth]{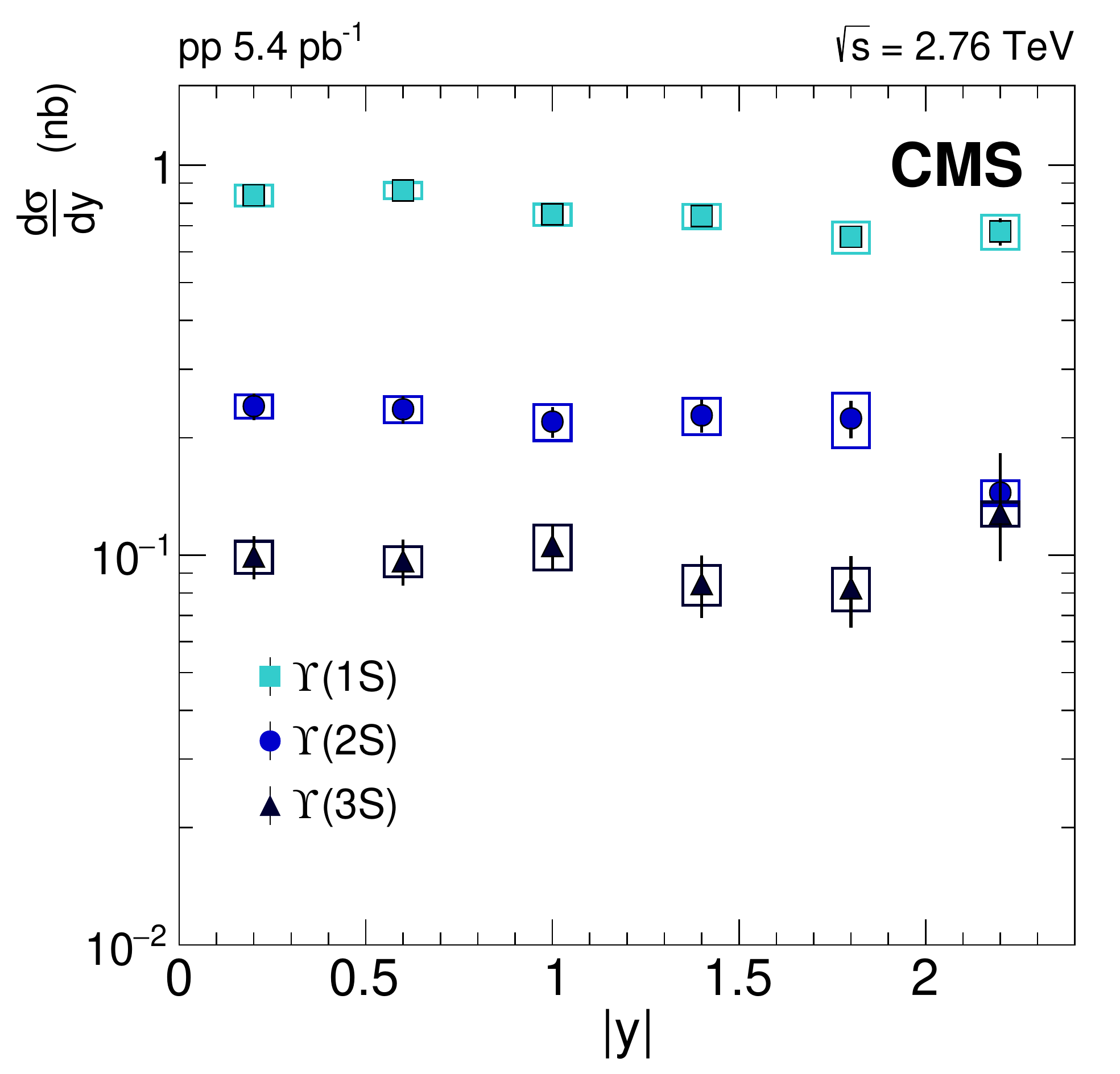}
\includegraphics[width=0.49\textwidth]{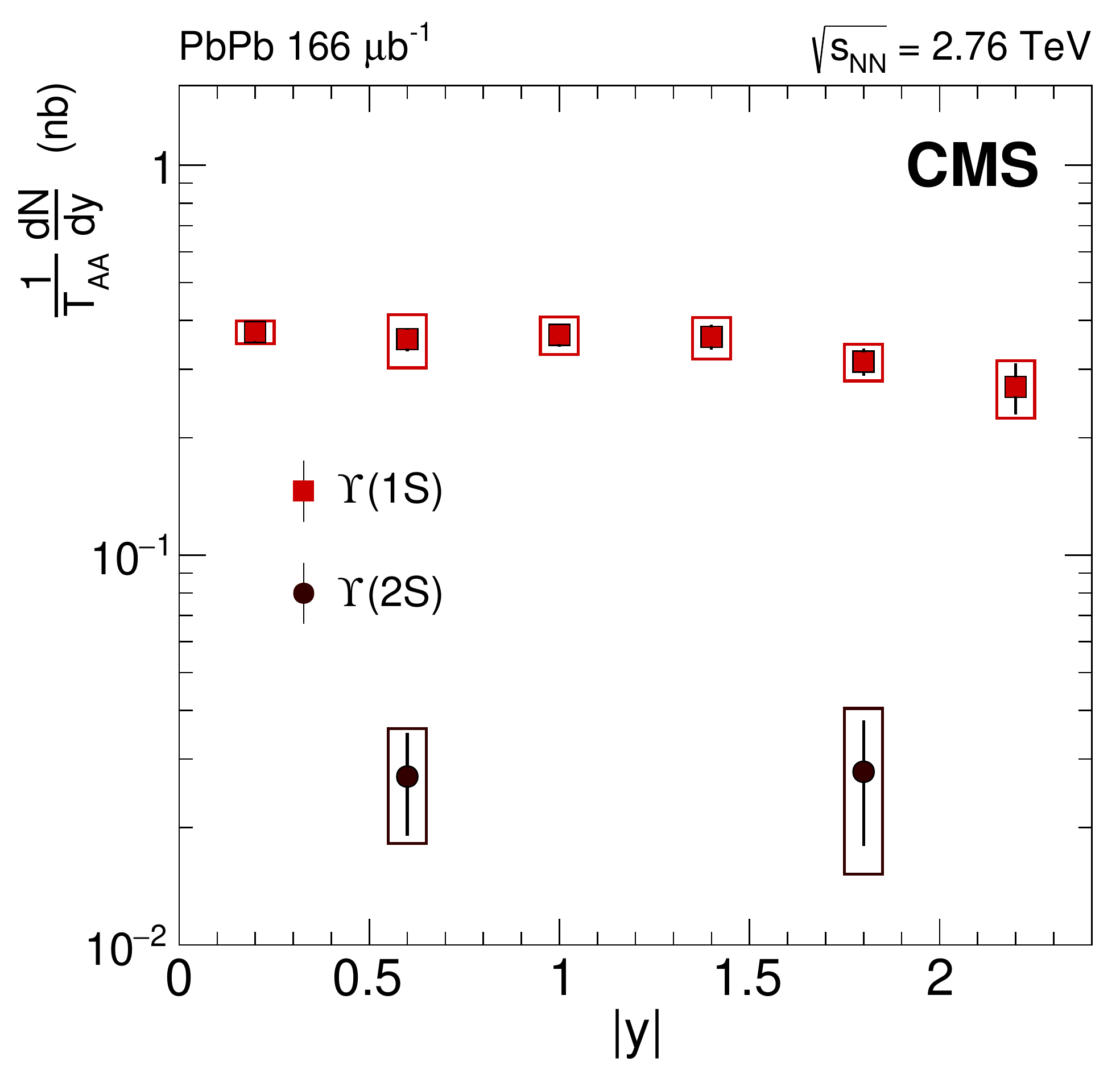}
\caption{Differential cross section for \PgU\ states as a function of their rapidity and integrated over transverse momentum in pp (\cmsLeft) and PbPb (\cmsRight) collisions. The PbPb results are integrated over centrality and divided by the number of elementary NN collisions. Statistical (systematic) uncertainties are displayed as error bars (boxes). Global relative uncertainties of 3.7\% (pp) and 6.5\% (PbPb) are not displayed.
}
\label{fig:CrossSection_ppAA_rap} 
\end{figure}

Figures~\ref{fig:CrossSection_ppAA_pt} and \ref{fig:CrossSection_ppAA_rap} show the differential cross sections as functions of \pt (per unit of rapidity) and $\abs{y}$, respectively, in pp (left) and PbPb (right) collisions. Measured yields are corrected for the acceptance and efficiency, then divided by the width of the bin in consideration. To put the pp and PbPb data on a comparable scale, the corrected yields are normalized by the measured integrated luminosity in pp collisions, and by the product of the number of corresponding MB events and the centrality-integrated \taa\ value for PbPb collisions. The statistical uncertainties in pp collisions allow a measurement for the three states using the same binning: five bins in \pt\ with edges at 0, 2.5, 5.0, 8.0, 12.0, and 20.0\GeVc, and six equal bins in $\abs{y}$ from 0 to 2.4. In PbPb collisions, that same binning can be used for the \PgUa\ analysis, but wider bins are necessary in the \PgUb\ case: three bins in \pt\ with edges at 0, 5, 12 and 20\GeVc, and two bins in $y$. The \PgUc\ state is not observed in PbPb collisions, and an upper limit is obtained for the \pt-, $y$- and centrality-integrated yield. The corresponding global (fully correlated) uncertainties (not shown in the plots) include the uncertainty due to the integrated luminosity in pp data, the uncertainties due to \taa and the number of MB events in PbPb data, and the uncertainty in the tracking efficiency in both cases.

\subsection{Nuclear modification factors}

\begin{figure}[hbtp]
\centering
\includegraphics[width=0.49\textwidth]{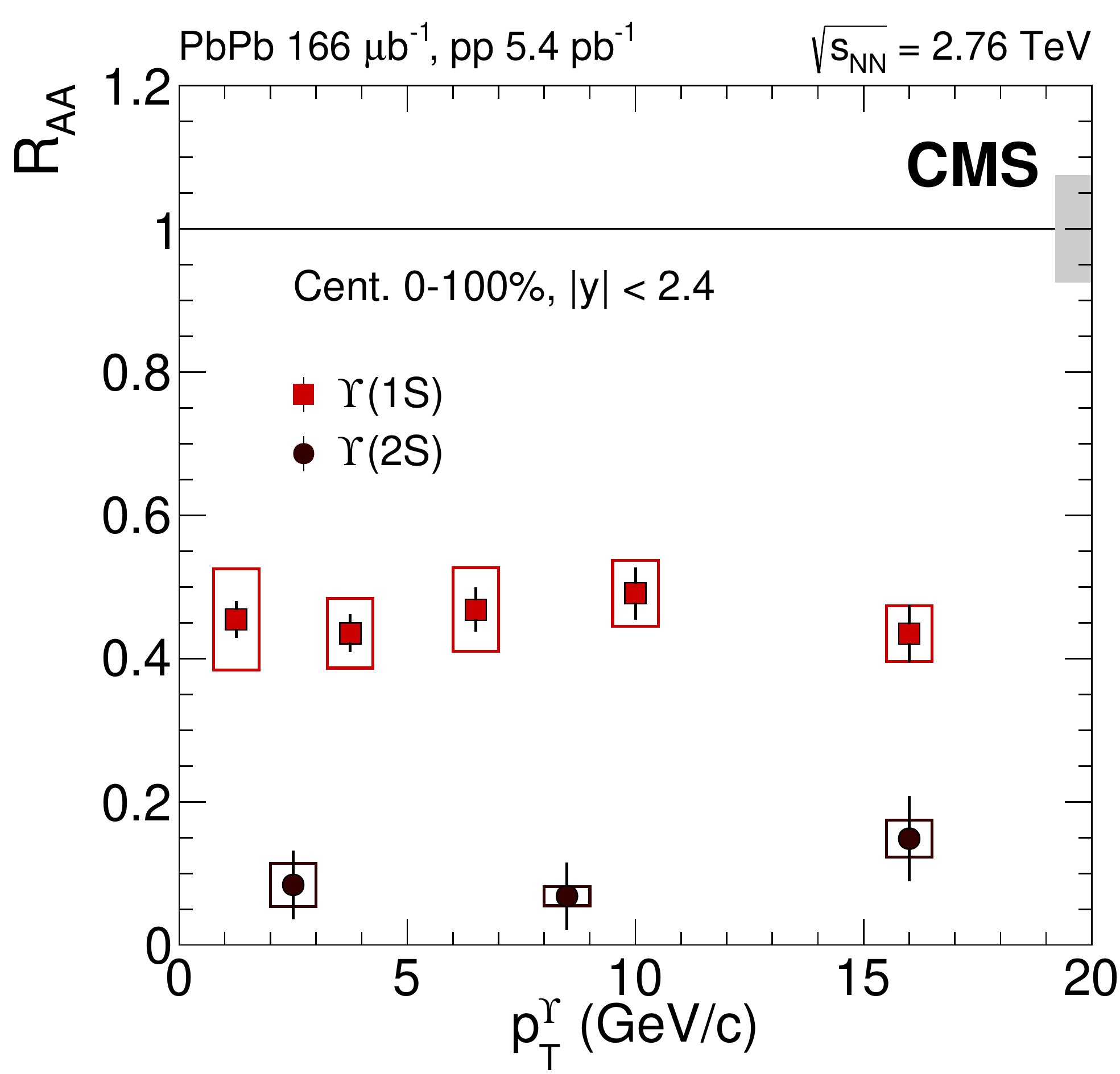}
\includegraphics[width=0.49\textwidth]{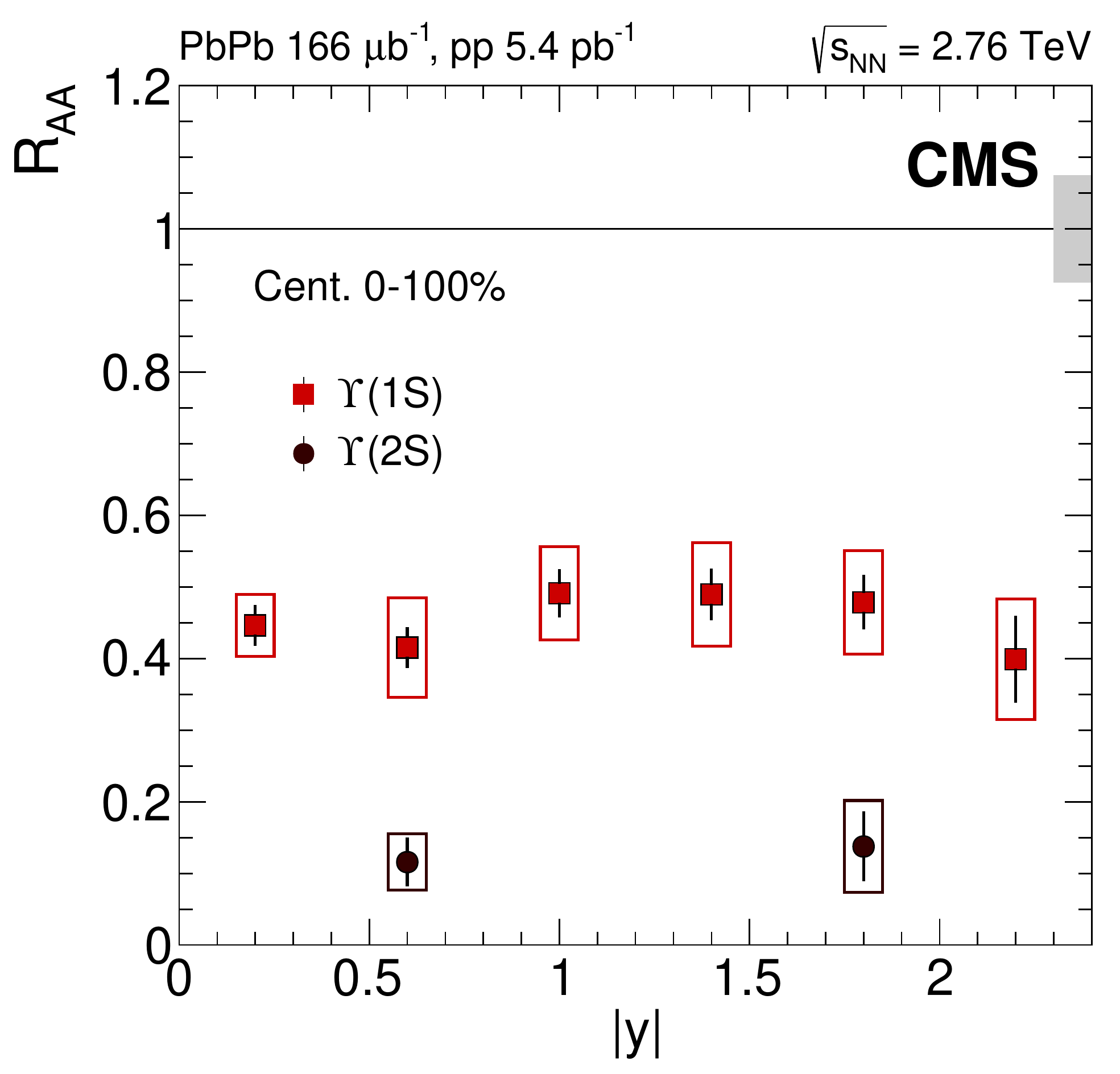}
\caption{Nuclear modification factor for \PgUa\ and \PgUb\ states in PbPb collisions as a function of \pt (\cmsLeft) and $\abs{y}$ (\cmsRight). Statistical (systematic) uncertainties are displayed as error bars (boxes), while the global (fully correlated) uncertainty (7.5\%) is displayed as a grey box at unity. }
\label{fig:Raa_kin}
\end{figure}

Nuclear modification factors, \raa, obtained by dividing the PbPb yields by the product of the \taa\ values and the pp cross sections, are shown on Fig. \ref{fig:Raa_kin} as a function of the \PgU\ meson \pt (left) and $\abs{y}$ (right). The global (fully correlated) uncertainty here includes the uncertainties in tracking efficiency, the integrated luminosity of the pp data, the number of MB PbPb events, and the centrality-integrated \taa\ value. The \raa results show a suppression of a factor of ${\approx}2$ and 8 for \PgUa\ and \PgUb\ states, respectively. No pronounced dependence on the \PgU\ meson kinematics is observed, the values being constant within uncertainties as a function of both \pt and $y$. 

\begin{figure}[hbtp]
\centering
\includegraphics[width=\cmsFigWidth]{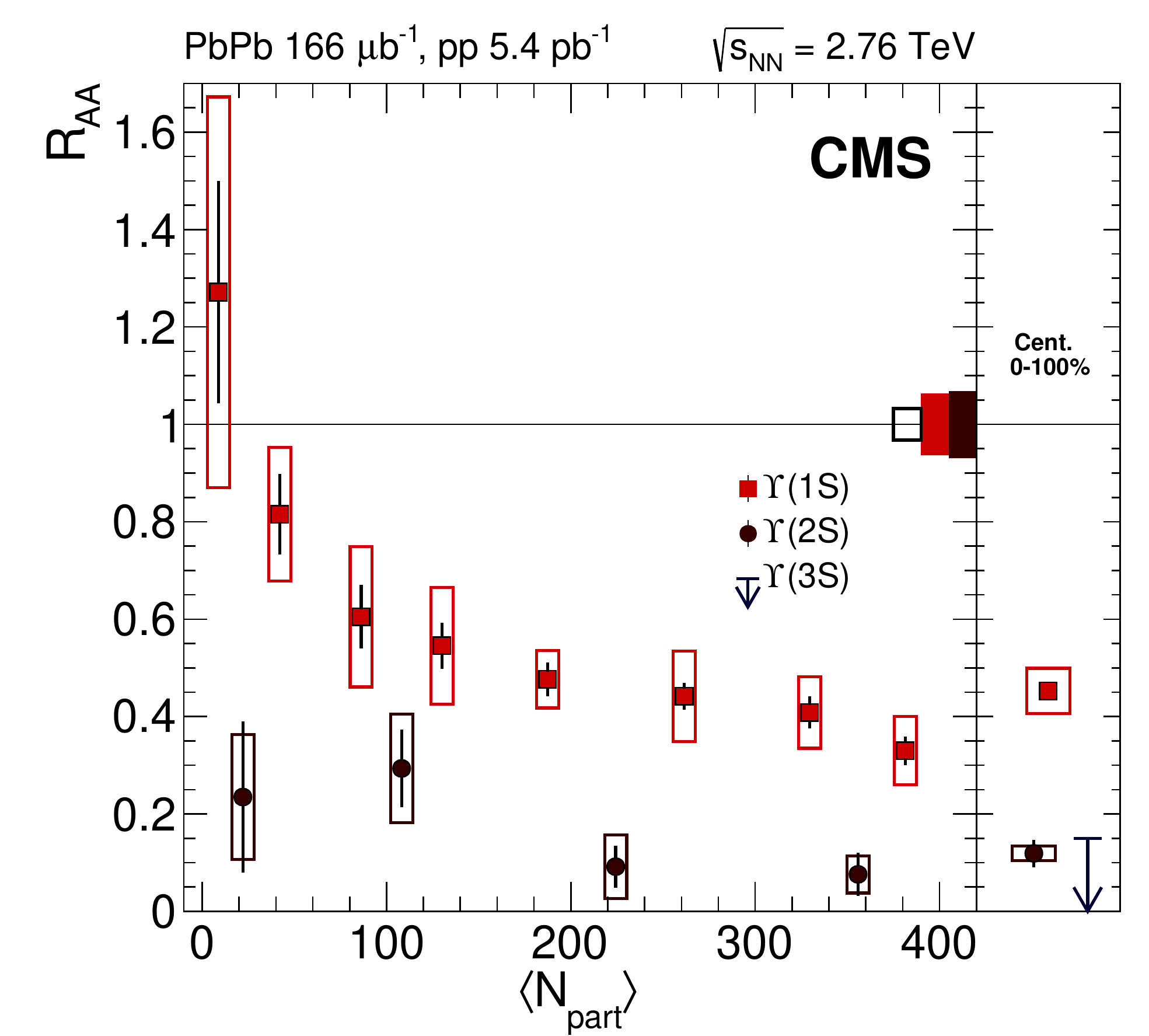}
\caption{Nuclear modification factors for \PgUa\ and \PgUb\ meson production in PbPb collisions, as a function of centrality, displayed as the average number of participating nucleons. The upper limit derived on the nuclear modification factor for \PgUc\ is represented with an arrow in the centrality integrated panel at the far right. Statistical (systematic) uncertainties are displayed as error bars (boxes), while the global (fully correlated) uncertainties from the PbPb data (3.2\%) or from the pp reference (6.3 and 6.9\% for \PgUa\ and \PgUb\ states, respectively) are displayed at unity as empty, filled red, and filled black boxes, respectively.}
\label{fig:Raa_cent}
\end{figure}

Figure~\ref{fig:Raa_cent} shows \raa as a function of centrality, displayed as the average number of participating nucleons, $\langle\npart\rangle$. The global (fully correlated) uncertainties come from the uncertainty in the pp cross sections (which differ for each \PgU\ state), the number of MB PbPb collisions and the PbPb tracking efficiency. The noticeable \PgUa\ centrality dependence, already observed in Ref.~\cite{Chatrchyan:2012lxa}, is mapped out with more precision. As discussed in Section~\ref{sec:centrality}, points are displayed at the~\npart value found by averaging over all MB events in each centrality class. In that respect, it should be noted that the large \PgUb\ suppression observed for the 50--100\% centrality range spans a wide range of \npart values, over which suppression could significantly change. The \raa values integrated over centrality for the three \PgU\ states are shown in the side panel of Fig.~\ref{fig:Raa_cent}. 
 
The lack of observation of the \PgUc\ state in PbPb data provides an upper limit on \raa, using the Feldman--Cousins prescription~\cite{Feldman:1997qc}. The centrality-integrated \raa values for the three states are:
\begin{align*}
\raa (\PgUa) & =  0.453 \pm 0.014 \pm 0.046; \\ 
\raa (\PgUb) & =  0.119 \pm 0.028 \pm 0.015; \\ 
\raa (\PgUc) & <  0.145\ \text{at a 95\% confidence level, }
\end{align*}
with the first and second uncertainties being one standard deviation statistical and systematic, respectively. 

These observations are consistent with a sequential melting scenario for the \PgU\ states, as described in Refs.~\cite{Emerick:2011xu,Krouppa:2016jcl,Hoelck:2016tqf}. 
These models, which attribute most of the suppression to in-medium melting, do not predict a strong dependence of \raa\ on rapidity or transverse momentum. Cold nuclear matter effects such as PDF modifications and energy loss also do not exhibit such dependences, and their overall impact on \PgU\ states is much smaller than the observed suppression~\cite{Arleo:2014oha}. In contrast, quarkonium regeneration should depend significantly on \pt, but it is predicted to be small for bottom quarks~\cite{Emerick:2011xu}. The sequential suppression by comoving particles computed in Ref.~\cite{Ferreiro:2014bia} reproduces the \PgU\ suppression centrality pattern, but any dependence on either \pt\ or $y$ remains to be assessed.

\section{Summary}

The \PgUa, \PgUb, and \PgUc\ yields have been measured in PbPb and pp collisions at $\sqrtsnn = 2.76$\TeV with the CMS detector, using integrated luminosities of 166\mubinv and 5.4\pbinv, respectively. For the first time, differential production cross sections are derived for individual \PgU\ states as functions of their rapidity and transverse momentum in heavy ion collisions. The \PgUa\ and \PgUb\ states are suppressed in PbPb relative to pp collisions scaled by the number of nucleon-nucleon collisions, by factors of ${\approx}2$ and 8, respectively, while the absence of a significant \PgUc\ signal corresponds to a suppression by a factor larger than ${\approx}7$ at a 95\% confidence level. While a strong centrality dependence of the suppression is found for the \PgUa\ and \PgUb\ states, no clear dependence is observed as a function of either transverse momentum or rapidity. The level of suppression measured in this analysis is compatible with theoretical models of a sequential melting of quarkonium states in a hot medium.

\begin{acknowledgments}
We congratulate our colleagues in the CERN accelerator departments for
the excellent performance of the LHC and thank the technical and
administrative staffs at CERN and at other CMS institutes for their
contributions to the success of the CMS effort. In addition, we
gratefully acknowledge the computing centres and personnel of the
Worldwide LHC Computing Grid for delivering so effectively the
computing infrastructure essential to our analyses. Finally, we
acknowledge the enduring support for the construction and operation of
the LHC and the CMS detector provided by the following funding
agencies: BMWFW and FWF (Austria); FNRS and FWO (Belgium); CNPq,
CAPES, FAPERJ, and FAPESP (Brazil); MES (Bulgaria); CERN; CAS, MoST,
and NSFC (China); COLCIENCIAS (Colombia); MSES and CSF (Croatia); RPF
(Cyprus); SENESCYT (Ecuador); MoER, ERC IUT and ERDF (Estonia);
Academy of Finland, MEC, and HIP (Finland); CEA and CNRS/IN2P3
(France); BMBF, DFG, and HGF (Germany); GSRT (Greece); OTKA and NIH
(Hungary); DAE and DST (India); IPM (Iran); SFI (Ireland); INFN
(Italy); MSIP and NRF (Republic of Korea); LAS (Lithuania); MOE and UM
(Malaysia); BUAP, CINVESTAV, CONACYT, LNS, SEP, and UASLP-FAI
(Mexico); MBIE (New Zealand); PAEC (Pakistan); MSHE and NSC (Poland);
FCT (Portugal); JINR (Dubna); MON, RosAtom, RAS and RFBR (Russia);
MESTD (Serbia); SEIDI and CPAN (Spain); Swiss Funding Agencies
(Switzerland); MST (Taipei); ThEPCenter, IPST, STAR and NSTDA
(Thailand); TUBITAK and TAEK (Turkey); NASU and SFFR (Ukraine); STFC
(United Kingdom); DOE and NSF (USA). 

Individuals have received support from the Marie-Curie programme and
the European Research Council and EPLANET (European Union); the
Leventis Foundation; the A. P. Sloan Foundation; the Alexander von
Humboldt Foundation; the Belgian Federal Science Policy Office; the
Fonds pour la Formation \`a la Recherche dans l'Industrie et dans
l'Agriculture (FRIA-Belgium); the Agentschap voor Innovatie door
Wetenschap en Technologie (IWT-Belgium); the Ministry of Education,
Youth and Sports (MEYS) of the Czech Republic; the Council of Science
and Industrial Research, India; the HOMING PLUS programme of the
Foundation for Polish Science, cofinanced from European Union,
Regional Development Fund, the Mobility Plus programme of the Ministry
of Science and Higher Education, the National Science Center (Poland),
contracts Harmonia 2014/14/M/ST2/00428, Opus 2013/11/B/ST2/04202,
2014/13/B/ST2/02543 and 2014/15/B/ST2/03998, Sonata-bis
2012/07/E/ST2/01406; the Thalis and Aristeia programmes cofinanced by
EU-ESF and the Greek NSRF; the National Priorities Research Program by
Qatar National Research Fund; the Programa Clar\'in-COFUND del
Principado de Asturias; the Rachadapisek Sompot Fund for Postdoctoral
Fellowship, Chulalongkorn University and the Chulalongkorn Academic
into Its 2nd Century Project Advancement Project (Thailand); and the
Welch Foundation, contract C-1845.
\end{acknowledgments}
\bibliography{auto_generated}

\cleardoublepage \appendix\section{The CMS Collaboration \label{app:collab}}\begin{sloppypar}\hyphenpenalty=5000\widowpenalty=500\clubpenalty=5000\textbf{Yerevan Physics Institute,  Yerevan,  Armenia}\\*[0pt]
V.~Khachatryan, A.M.~Sirunyan, A.~Tumasyan
\vskip\cmsinstskip
\textbf{Institut f\"{u}r Hochenergiephysik,  Wien,  Austria}\\*[0pt]
W.~Adam, E.~Asilar, T.~Bergauer, J.~Brandstetter, E.~Brondolin, M.~Dragicevic, J.~Er\"{o}, M.~Flechl, M.~Friedl, R.~Fr\"{u}hwirth\cmsAuthorMark{1}, V.M.~Ghete, C.~Hartl, N.~H\"{o}rmann, J.~Hrubec, M.~Jeitler\cmsAuthorMark{1}, A.~K\"{o}nig, I.~Kr\"{a}tschmer, D.~Liko, T.~Matsushita, I.~Mikulec, D.~Rabady, N.~Rad, B.~Rahbaran, H.~Rohringer, J.~Schieck\cmsAuthorMark{1}, J.~Strauss, W.~Waltenberger, C.-E.~Wulz\cmsAuthorMark{1}
% ...
\vskip\cmsinstskip
% ...
1:~~Also at Vienna University of Technology, Vienna, Austria\\
% ...

\end{sloppypar}
\end{document}